\documentclass[10pt]{iopart}

\usepackage{iopams}

\expandafter\let\csname equation*\endcsname\relax
\expandafter\let\csname endequation*\endcsname\relax

\usepackage[english]{babel}
\usepackage{amsmath}
\usepackage{color}
\usepackage{epsfig}
\usepackage{graphicx}
\usepackage{bm}
\usepackage{times}
\usepackage{subfigure}
\usepackage{color}
\usepackage{leftidx}

\newcommand{\rev}[1]{#1}

\global\long\def\tr{\mathop{\mathrm{tr}}\limits}
\global\long\def\trb{\mathop{\mathrm{tr}_{1,N}}\limits}
\global\long\def\trn#1{\mathop{\mathrm{tr}_{#1}}\limits}

\global\long\def\bra#1{\left\langle #1\right|}
\global\long\def\ket#1{\left|#1\right\rangle }

\global\long\def\ketbra#1#2{\ket{#1}\bra{#2}}

\global\long\def\al{\alpha}
\global\long\def\be{\beta}

\global\long\def\De{\Delta}
\global\long\def\Ga{\Gamma}
\global\long\def\th{\theta}

\global\long\def\la{\lambda}
\global\long\def\ka{\kappa}
\global\long\def\si{\sigma}
\global\long\def\vfi{\varphi}

\begin{document}

\title[Spin-helix states in the $XXZ$ spin chain with strong boundary
dissipation]{Spin-helix states in the $XXZ$ spin chain with strong
  dissipation}

\author{Vladislav Popkov} \address{ HISKP, University of Bonn,
  Nussallee 14-16, 53115 Bonn, Germany.}  \address{ Centro
  Interdipartimentale per lo studio di Dinamiche Complesse,
  Universit\`a di Firenze, via G.  Sansone 1, 50019 Sesto Fiorentino,
  Italy } \ead{vladipopkov@gmail.com} \author{Johannes Schmidt}
\address{Institut f\"{u}r Teoretische Physik, Universit\"{a}t zu
  K\"{o}ln, Z\"ulpicher str. 77, K\"{o}ln, Germany.}
\ead{schmidt@thp.uni-koeln.de} \author{Carlo Presilla}
\address{Dipartimento di Fisica, Sapienza Universit\`a di Roma,
  Piazzale Aldo Moro 2, Roma 00185, Italy} \address{Istituto Nazionale
  di Fisica Nucleare, Sezione di Roma 1, Roma 00185, Italy}
\ead{carlo.presilla@roma1.infn.it} \vspace{10pt}

\begin{indented}
\item[]\today
\end{indented}

% \thanks{E-mail: vpopkov@uni-koeln.de} \affiliation{Institut f\"{u}r
% Theoretische Physik, Universit\"{a}t zu K\"{o}ln, Z\"uelicher
% str. 77, K\"{o}ln, Germany.}

% \author{Carlo Presilla} \affiliation{Dipartimento di Fisica,
% Sapienza Universit\`a di Roma, Piazzale Aldo Moro 2, Roma 00185,
% Italy} \affiliation{Istituto Nazionale di Fisica Nucleare, Sezione
% di Roma 1, Roma 00185, Italy}

\begin{abstract}
  We investigate the nonequilibrium steady state (NESS) in an open
  quantum XXZ chain \rev{attached at the ends to polarization baths
    with unequal polarizations.}  Using the general theory developed
  in \cite{ZenoGeneralTheory}, we show that in the critical $XXZ$
  $|\Delta|<1$ easy plane case, the steady current in large systems
  under strong driving shows resonance-like behaviour, by an
  infinitesimal change of the spin chain anisotropy or other
  parameters.  Alternatively, by fine tuning the system parameters and
  varying the boundary dissipation strength, we observe a change of
  the NESS current from diffusive (of order $1/N$, for small
  dissipation strength) to ballistic regime (of order 1, for large
  dissipation strength).  This drastic change results from an
  accompanying structural change of the NESS, which becomes a pure
  spin-helix state characterized by a winding number which is
  proportional to the system size.  We calculate the critical
  dissipation strength needed to observe this surprising effect.
\end{abstract}
% \pacs{75.10.Pq, 03.65.Yz, 05.60.Gg, 05.70.Ln} \pacs{75.10.Pq,
% 03.65.Yz, 05.60.Gg, 05.70.Ln}

% \maketitle

\begin{figure}[tbp]
  \centerline{\includegraphics[width=8cm,height=6cm,clip]%
    {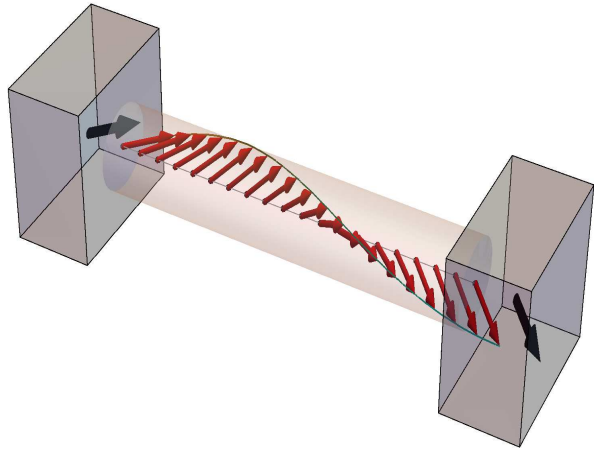}}
  \caption{Schematic setup of a chain of spins attached to two fully
    polarizing boundary reservoirs. The chain has $N=20$ spins and we
    chose boundary conditions $\theta_L=\theta_R=0.4$, $\varphi_L=0$,
    $\varphi_R=4$.}
  \label{Fig_setup}%
\end{figure}

% \section{Introduction}
The $XXZ$ Heisenberg spin chain is a paradigmatic model in statistical
mechanics. Its remarkable properties are long known in thermodynamic
equilibrium \cite{Gaudin83,QISM93}. Recently, it was shown that also
in a nonequilibrium setting, under a non-coherent boundary driving,
the $XXZ$ chain retains many remarkable properties.  An interesting
strongly nonequilibrium setup of the problem occurs when a coherent
evolution in the bulk is accompanied by a non-coherent local boundary
driving, which tends to polarize the boundary spins along two
different directions.  If the boundary baths do not match, the system
experiences a gradient of magnetization which leads to nonzero
currents, even in the steady state. A schematic setup of the model is
shown in Fig.~\ref{Fig_setup}.  Note that the alignment of the
boundary spins to the respective baths cannot be made perfect due to
quantum fluctuations, except for the so-called Zeno limit, when the
boundary dissipation is infinitely strong.  An interplay between
coherent bulk effects and incoherent boundary couplings results in the
nontrivial scaling properties of the nonequilibrium steady state
(NESS) characteristics (the currents, density profiles, many-point
correlations, etc.), which can be distinguished as different phases of
criticality of the $XXZ$ model
\cite{2011ZnidaricA,2011ZnidaricB,ProsenExact2011}.

Precise structure of the NESS for large systems and arbitrary
\rev{mismatch of boundary polarizations} is out of reach because the
complexity of the problem grows exponentially with the system size
$N$.  From the general setup of the problem one would naively expect
the NESS magnetization profile to interpolate between the left and
right boundary as depicted in Fig.~\ref{Fig_setup}.  A few solvable
cases, for which the NESS can be found analytically,
\cite{2015ProsenReview} suggest that the system properties essentially
depend on the phases of criticality of the $XXZ$ model, characterized
by the value of the spin exchange anisotropy $\Delta$, while the NESS
within each phase separately varies regularly and smoothly.

In the present communication we demonstrate that, contrary to the
expectations, the regular analytic behaviour of the NESS breaks down
in a seemingly innocent and natural situation when the boundary
driving is combined with an arbitrary spin-exchange anisotropy.  We
find that for a set of fine-tuned values of the anisotropy, various
characteristics of the NESS, e.g., the magnetization current, may
change dramatically, by orders of magnitude, and from monotonic
behaviour to strongly nonmonotonic, provided that the dissipative
strength $\Gamma$ becomes sufficiently large. For these special
anisotropy values, and in their proximity as well, a structural
transition in the NESS occurs, from a spatially smooth local
magnetization profile interpolating between the boundary baths (small
$k$ in the Fourier space), to a rigid quasi-periodic structure of
spins corresponding to large $k$ values, arranged in a helix.  Such a
drastic structural transition naturally entails a singular behaviour
of the NESS. Remarkably, spin-helix state is a pure state, which is
rather unusual for a many-body interacting quantum system
dissipatively coupled to an external bath.  Detuning the anisotropy or
lowering the dissipation strength below a threshold value makes the
spin-helix structure to relax back to a smooth profile.  The set of
critical anisotropies, at which the structural transitions to
spin-helix state occur, becomes dense on the segment $[-1,1]$ in the
limit of large system size $N$.

The plan of the paper is as follows. We introduce the model and
various properties of interest in Sec.~\ref{sec::Model}.  In
Sec.~\ref{sec::condition for NESS purity} we review the conditions
under which the pure NESS is achieved in the Zeno limit. In
Sec.~\ref{sec::Convergence to the spin-helix state} the convergence to
an atypical NESS for finite dissipative strength is quantified, while
in Sec.~\ref{sec::divergences of Gammach} we characterize the points
where this convergence fails. We discuss two possible experimental
scenarios in Sec.~\ref{sec::Experimental scenarios} and, finally, in
Sec.~\ref{sec::Conclusions} we draw our conclusions.

\section{Model}
\label{sec::Model}

We consider an open $XXZ$ chain coupled dissipatively to boundary
reservoirs, described via the Lindblad master
equation~\cite{Petruccione,PlenioJumps,ClarkPriorMPA2010}
\begin{equation}
  \frac{\partial\rho}{\partial t}=
  -\frac{i}{\hbar}\left[  H,\rho\right]
  +\sum_\alpha L_\alpha \rho L^\dag_\alpha- \frac{1}{2}
  \left(  L_\alpha^\dag L_\alpha \rho + \rho L_\alpha^\dag L_\alpha \right),
  \label{LME}%
\end{equation}
where $H$ is the spin $1/2$ Heisenberg Hamiltonian with a partial
anisotropy along the $Z$-axis
\begin{align}
  H_{XXZ}&= \sum_{j=1}^{N-1} h_{j,j+1}^{XXZ}= \sum_{j=1}^{N-1} J
  \left( \sigma_{j}^{x}\sigma_{j+1}^{x}+
    \sigma_{j}^{y}\sigma_{j+1}^{y}+ \Delta \left(
      \sigma_{j}^{z}\sigma_{j+1}^{z}- I \right)
  \right). \label{DefXXZ}%
\end{align}
The parameter $\Delta$ describes the $Z$-anisotropy.  The Lindblad
operators are chosen so as to target completely polarized states of
the leftmost and rightmost spins (spins number 1 and number $N$,
respectively). We parametrize the targeted boundary polarizations by
polar and azimuthal angles $\theta_L,\varphi_L$ on the left end of the
chain and $\theta_R,\varphi_R$ on its right end.  We consider only two
Lindblad operators, $L_1, L_2$, the first one being
\begin{align}
  L_1= \frac{\sqrt{\Gamma}}{2} \left( -\sin \theta_L \sigma_1^z
    +(1+\cos \theta_L )e^{-i \varphi_L} \sigma_1^{+} + (-1+\cos
    \theta_L )e^{i \varphi_L} \sigma_1^{-}\right),
  \label{DefL1}%
\end{align}
where $\sigma^\alpha$, $\al=x,y,z$, are Pauli matrices, lower indices
denote the embeddings in the physical space, and
$\sigma^{\pm}=(\sigma^x \pm i \sigma^y)/2$. The second Lindblad
operator, $L_2$, is obtained from Eq.~(\ref{DefL1}) by the
substitutions $\sigma_1^\alpha \rightarrow \sigma_N^\alpha$, $\theta_L
\rightarrow \theta_R$, $\varphi_L \rightarrow \varphi_R$.  It can be
straightforwardly verified, that the pure one-site state
$\rho_L=\ket{\psi_1}\bra{\psi_1}$, with $\bra{\psi_1}= \langle
\cos(\theta_L/2) e^{-i \varphi_L/2}, \sin(\theta_L/2) e^{i
  \varphi_L/2} |$, is a dark state of $L_1$, i.e., $L_1
\ket{\psi_1}=0$. In the absence of the coherent evolution term in
Eq.~(\ref{LME}), the left boundary spin relaxes to a state $\rho_L=
\ket{\psi_1}\bra{\psi_1}=\frac{1}{2} \vec{l}_L\vec{\sigma}_1$, with
$\vec{l}_L=(\sin \theta_L \cos \varphi_L, \sin \theta_L \sin
\varphi_L, \cos\theta_L)$, with a characteristic time $\tau =
\Gamma^{-1}$. An analogous statement holds for the rightmost spin,
which (in the absence of the coherent evolution) gets polarized along
the direction $\vec{l}_R=(\sin \theta_R \cos \varphi_R, \sin \theta_R
\sin \varphi_R, \cos\theta_R)$. A possible experimental protocol of
repeating interactions leading to the density matrix evolution
(\ref{LME}) (\ref{DefL1}) is discussed in \cite{Landi}.  It is clear
that any non fully matching boundary conditions, $(\theta_L,\varphi_L)
\neq (\theta_R,\varphi_R)$, introduces a boundary \rev{mismatch}, and
results in steady currents flowing through the chain. In particular,
due to the spin-exchange anisotropy in $XY$-plane, the $Z$-component
of the magnetization current $j^z$ is locally conserved.

In the following solvable cases, the NESS, namely, the
time-independent solution of Eq.~(\ref{LME}), is known analytically:

\textit{Collinear boundary driving along the anisotropy axis
  $\varphi_L=\varphi_R=0, \theta_L=0, \theta_R=\pi$ }
\cite{ProsenExact2011}. The steady magnetization current is ballistic
in the critical regime $|\Delta| <1$, is exponentially small in the
Ising-like case $|\Delta| >1$ and is subdiffusive in the isotropic
case $\Delta =1$.  For large $\Gamma$, the $Z$-component of the
magnetization current $j^z$ vanishes due to quantum Zeno effect.

\textit{Non-collinear $XY$-plane boundary driving
  $\varphi_L=0,\varphi_R=\Phi, \theta_L= \theta_R=\pi/2$ and isotropic
  Heisenberg model $\Delta=1$} \cite{MPA,MPA-PRE2013}. In the
isotropic case, all components of the magnetization current are
conserved. The components $j^x, j^y$ are subdiffusive and decrease for
large $\Gamma$, due to quantum Zeno effect, while $j^z$ monotonically
increases with $\Gamma$.  The NESS-dependence on $\Gamma$ is regular
and piecewise monotonic.

\textit{Non-collinear strong $XY$-plane boundary driving and
  fine-tuned anisotropy $\Delta$}. It was suggested in
\cite{PhysRevA.93.022111}, that for sufficiently strong dissipative
coupling, the NESS becomes arbitrarily close to a pure state, which we
shall call spin-helix state (SHS), in analogy to states appearing in
two-dimensional electron systems with spin-orbit coupling
\cite{PhysRevLett.97.236601,Koralek2009,Winkler03}, $\lim_{\Gamma
  \rightarrow \infty} \rho_\mathrm{NESS}(\Gamma)= \ket{
  \Psi_\mathrm{SHS}}\bra{ \Psi_\mathrm{SHS}} $, where
\begin{equation}
  \ket{ \Psi_\mathrm{SHS}} = 2^{-\frac{N}{2}} \bigotimes_{k=1}^N
  \left(
    \begin{array}{c}
      e^{-\frac{i}{2}\varphi (k-1)}
      \\
      e^{\frac{i}{2}\varphi (k-1)}
    \end{array}
  \right),
  \label{XYplanetargetedstate}%
\end{equation}
provided the states of the boundary spins match the boundary driving,
namely, $\theta_L=\theta_R=\pi/2$, $\varphi_L=0, \varphi_R \equiv
(N-1) \varphi = \Phi$, and the anisotropy $\Delta$ obeys
\begin{equation} \Delta = \cos \varphi.
  \label{DeltaCritical0}%
\end{equation}
In the present situation, when the Lindblad operators (\ref{DefL1})
are targeting pure single-spin states, the spin-helix
states~(\ref{XYplanetargetedstate}) are obtained in an ideal Zeno
regime $\Gamma\to\infty$.  Note, however, that it is also possible to
generate the same spin-helix states for finite dissipative strengths
$\Gamma$, if fine-tuned mixed single-spin states at the boundaries are
dissipatively targeted~\cite{PopkovSchutzSHS}.

The spin-helix states~(\ref{XYplanetargetedstate}) are quite
remarkable in many respects. From the point of view of a dissipative
dynamics, the creation of a pure quantum state via a dissipative
action is a way to beat detrimental decoherence effects. From the
point of view of spintronics, the state (\ref{XYplanetargetedstate})
carries an anomalously high ballistic magnetization current of order
$1$, which is independent of the system size.

The existence of SHS in the Zeno limit at fine-tuned anisotropy
(\ref{DeltaCritical0}) was guessed in \cite{PhysRevA.93.022111} on the
base of a necessary criterion and explicit calculation of the NESS for
small system sizes.  Here we revisit and systematically treat the SHS
on the base of a general theory (which provides necessary and
sufficient criteria for SHS existence, and also a convergence
criterion) developed by us in \cite{ZenoGeneralTheory}.  We treat more
general SHS,
\begin{align}
  \ket{\Psi_\mathrm{SHS}} = \bigotimes_{k=1}^N \left(
    \begin{array}{c}
      \cos(\frac{\theta}{2})  e^{-\frac{i}{2}\varphi(k-1)}
      \\
      \sin(\frac{\theta}{2})  e^{\frac{i}{2}\varphi(k-1)}
    \end{array}
  \right),
  \label{XXZtargetedstate}%
\end{align}
which, as we shall see later, can be dissipatively generated in a
boundary driven $XXZ$ spin chain by tuning the boundary conditions and
the anisotropy.  The state (\ref{XXZtargetedstate}) describes a
precession, along the chain, of the local spin around the $Z$-axis,
forming a frozen spin wave structure, see Fig.~\ref{Fig_setup} for an
illustration, with constant twisting azimuthal angle difference
$\varphi$ between two neighbouring spins. This is evident if we
compute the expectation value of the local spin at site $n$
\begin{equation} \bra{ \Psi_\mathrm{SHS}}\vec{\sigma}_n \ket{
    \Psi_\mathrm{SHS}} = (\sin\theta \cos \varphi (n-1),\sin\theta
  \sin \varphi (n-1),\cos\theta).
  \label{ResLocalMagnetization}%
\end{equation}
The local spin orientations obtained in a chain with $N=20$ spins and
boundary conditions $\theta_L=\theta_R=0.4$, $\varphi_L=0$,
$\varphi_R=4$ are shown in Fig.~\ref{Fig_setup}.

Note that fixed boundary polarizations $\varphi_L=0, \varphi_R = \Phi$
match not just one spin-helix state (\ref{XXZtargetedstate}) with
$\varphi (N-1)= \Phi$, but also those with $\varphi (N-1)= \Phi+ 2
\pi$, $\varphi (N-1)= \Phi + 4 \pi$ etc., until $\varphi (N-1)= \Phi+
(N-2)2 \pi$. Thus, we shall also characterize a spin-helix state via a
winding number $m=\lfloor (N-1)\varphi/(2 \pi) \rfloor$, $\lfloor
\cdot \rfloor$ being the integer part.
\begin{figure}[tbp]
  \centerline{\includegraphics[width=8cm,height=6cm,clip]%
    {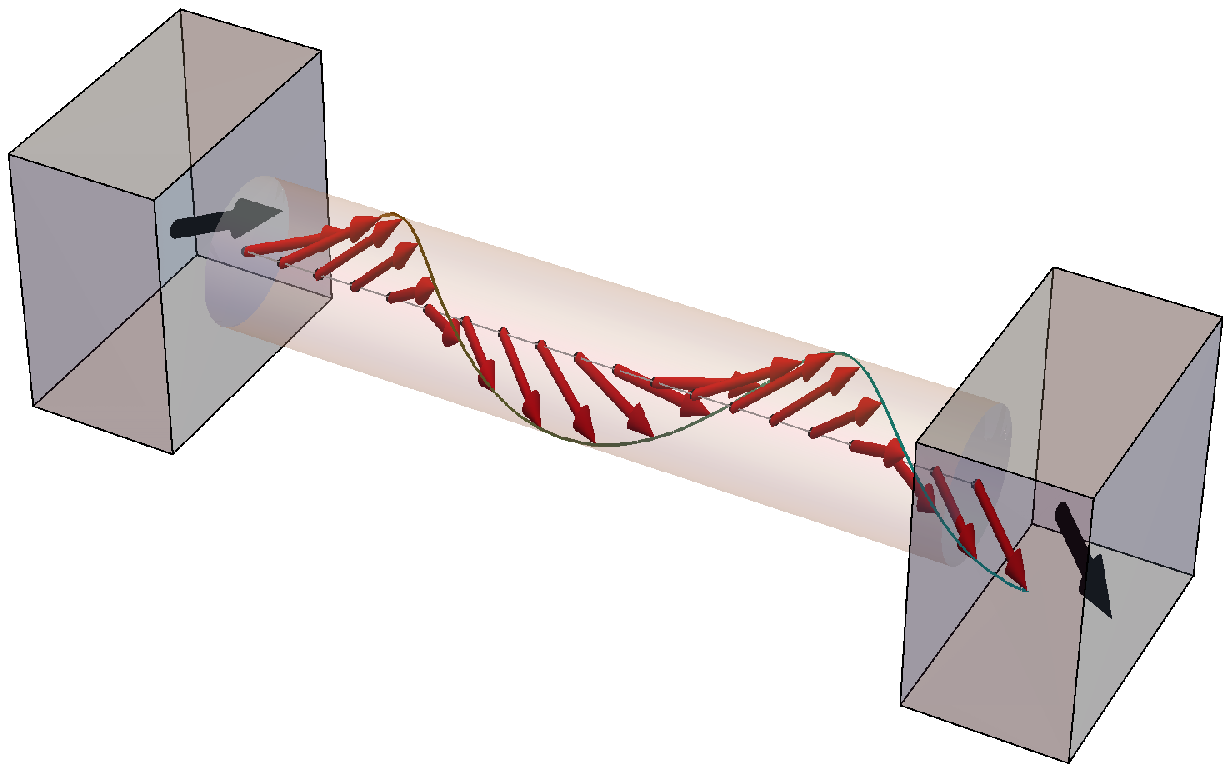}}
  \centerline{\includegraphics[width=8cm,height=6cm,clip]%
    {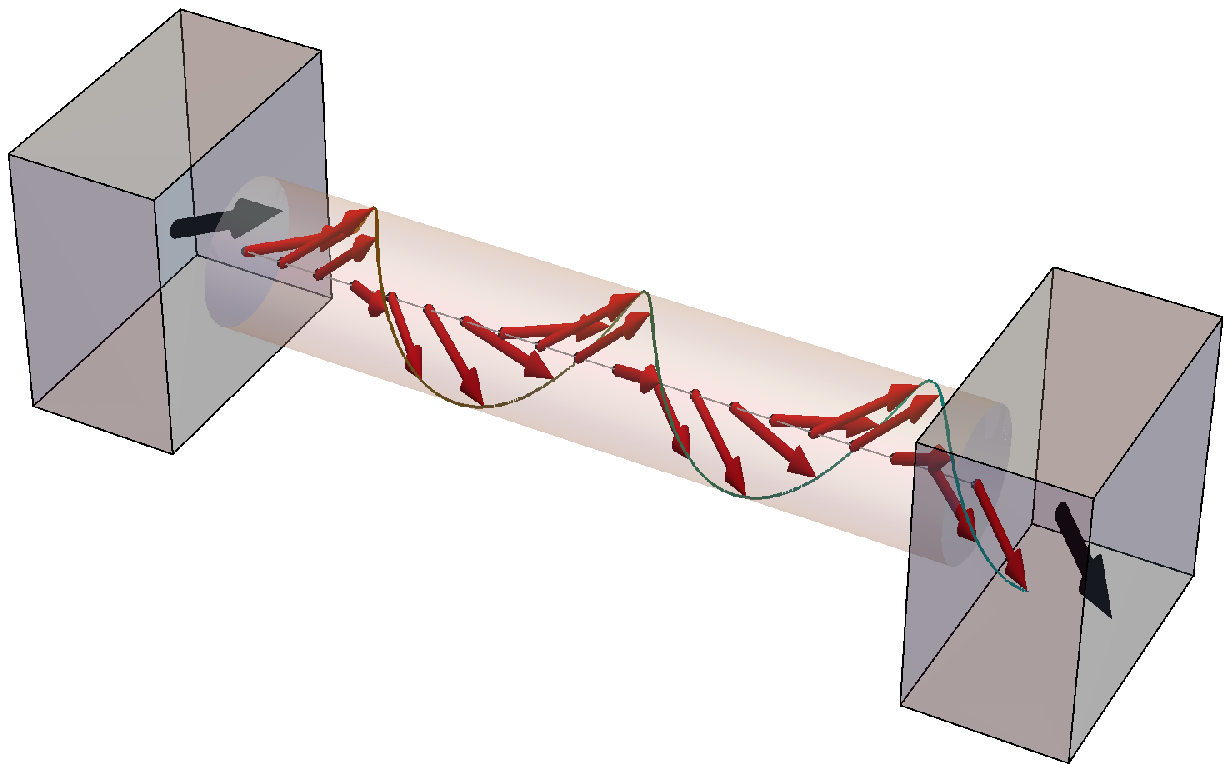}}
  \caption{spin-helix states as in Fig.~\ref{Fig_setup} but with
    winding numbers $m=1$ (top) and $m=2$ (bottom).}
  \label{Fig_setup_m_1_2}%
\end{figure}

We will see in the following that the spin-helix
states~(\ref{XXZtargetedstate}) constitute the points of
resonance-like behaviour of the NESS, which becomes visible at large
dissipation.  In doing that, we shall answer some basic questions.
How large must the dissipation strength be to reach the limiting
spin-helix state with a predefined accuracy? To which extent the
characteristics of the spin-helix state/states are \textit{atypical}
for given boundary mismatch? Can the resonance-like behaviour be
detected in other features of the NESS?  What happens if the system
gets larger and larger, and, eventually, we reach the thermodynamic
limit of infinitely long chains?

The characterization of several properties of the NESS prove to be
useful for our later considerations.  \newline (a) A measure of the
\rev{purity}. In fact, in driven Heisenberg spin chains with
polarization targeting operators, a NESS can become pure, e.g.,
$\rho_\mathrm{NESS}=\ketbra{\Psi_\mathrm{SHS}}{\Psi_\mathrm{SHS}}$,
where $\ket{\Psi_\mathrm{SHS}}$ is the spin-helix state
(\ref{XXZtargetedstate}), \textit{only} in the Zeno limit.  As a
criterion for \rev{purity} of a state $\rho$, we shall use both the
von Neumann entropy, $S_{VNE}(\rho)= - \tr(\rho \log_2 \rho)$, as well
as the alternative measure $\epsilon(\rho)= 1- \tr(\rho^2)$.  \newline
(b) Steady currents of magnetization and of energy. Being a
nonequilibrium steady state, the NESS is characterized by
non-vanishing steady currents. The magnetization (spin) current
operator in the $Z$-direction, $\hat{\jmath}_{n,n+1}$, is defined via
a lattice continuity equation $\frac
{d}{dt}\sigma_{n}^{z}=\hat{\jmath}_{n-1,n}-\hat{\jmath}_{n,n+1}$,
where
\begin{equation}
  \hat{\jmath}_{n,m} = J(\sigma_{n}^{x}\sigma_{m}^{y}
  -\sigma_{n}^{y}\sigma_{m}^{x}).
  \label{DefSpinCurr}%
\end{equation}
The energy current operator, $\hat{J}_{n}^{E}$, is defined analogously
by $\frac {d}{dt}h_{n,n+1}=\hat{J}_{n}^{E}-\hat{J}_{n+1}^{E}$, where
\begin{equation}
  \hat{J}_{n}^{E}=-\sigma_{n}^{z}\hat{\jmath}_{n-1,n+1}+\Delta(\hat{\jmath
  }_{n-1,n}\sigma_{n+1}^{z}+\sigma_{n-1}^{z}\hat{\jmath}_{n,n+1}).
  \label{DefEnergyCurr}%
\end{equation}
\newline (c) Finally, we need a cumulative function characterizing the
density profile $\sigma_{n}^{\alpha}$, which probes the helix
structure of the spins.  To this end, we introduce a generalized
structure factor (or, alternatively, a generalized discrete Fourier
Transform (GFT)), via
\begin{equation}
  \hat{f}_m(\Phi)= \frac{1}{M} \sum_{k=0}^{M-1} f_k e^{-i \varphi(m) k},
  \label{DefGeneralizedFourierTrasform}%
\end{equation}
where
\begin{equation}
  \varphi(m)= \frac{\Phi+2 \pi m}{M}, \qquad m=0,1,\dots, M-1.
\end{equation}
Here, $M+1=N$ is the chain length, $0 \leq \Phi < 2 \pi$ and $m$ is
the winding number. For $\Phi=0$,
Eq.~(\ref{DefGeneralizedFourierTrasform}) turns into the usual
discrete Fourier Transform.  The GFT shares similar properties with
the usual Fourier Transform, e.g., the Parseval identity has the usual
form
\begin{equation}
  \sum_{m=0}^{M-1} |\hat f_m(\Phi)|^2 = \frac{1}{M} \sum_{k=0}^{M-1} |f_k|^2.
  \label{IdentityParceval}%
\end{equation}
A convenient quantity to look at is the GFT
(\ref{DefGeneralizedFourierTrasform}) of the one-point observables
\begin{equation}
  f_{k-1}=\tr( (\sigma_k^x+i \sigma_k^y)\rho),\qquad k=1,2,\dots, N-1,
  \label{DefObservablesProfiles}%
\end{equation}
which play the role of the usual Fourier harmonics. Indeed, for a
spin-helix state with winding number $m_0$, we find $f_k= e^{i
  \varphi(m_0) k} $.
% and $\hat {f}_m(\Phi)=\delta_{m,m_0}$.

The above quantities (a)-(c) are easily calculated for the stationary
spin-helix state
$\rho_\mathrm{SHS}=\ketbra{\Psi_\mathrm{SHS}}{\Psi_\mathrm{SHS}}$,
where $\ket{\Psi_\mathrm{SHS}}$ is given by
Eq.~(\ref{XXZtargetedstate}) with $\varphi= \varphi(m_0)= (\Phi+2 \pi
m_0)/(N-1)$, yielding
\begin{align}
  S_{VNE}(\rho_\mathrm{SHS})&=\epsilon(\rho_\mathrm{SHS})=0 ,
  \\
  j^z(\rho_\mathrm{SHS})&=\tr( \hat{\jmath}_{n,n+1}
  \rho_\mathrm{SHS})= J \sin\theta \sin\varphi(m_0) ,
  \\
  J_E(\rho_\mathrm{SHS})&= \tr( \hat{J}_{n}^{E} \rho_\mathrm{SHS})= 0
  ,
  \\
  \hat {f}_m(\Phi) &= \sin\theta \delta_{m,m_0}, \qquad m=0,1,\dots,
  N-2.
  \label{SHSfourierCoeff}%
\end{align}

\begin{figure}[tbp]
  \centerline{\includegraphics[width=9cm,clip]%
    {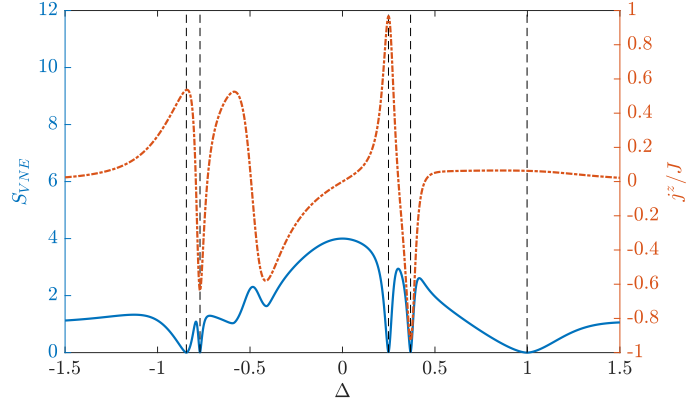}}
  \caption{von Neumann entropy (solid blue line\rev{,left vertical
      scale}) of the NESS and steady-state magnetization current
    (dot-dashed red line\rev{, right vertical scale}) versus the
    anisotropy $\Delta$ for $\Gamma=1000$.  The minima of $S_{VNE}$
    correspond to almost pure spin-helix states with different winding
    numbers.  The vertical dashed lines indicate the critical
    anisotropies of Eq.~(\ref{DeltaCritical}) obtained, from left to
    right, for $m=2,3,4,1,0$.  System parameters: $N=6$,
    $\theta=\pi/2$, $\Phi=\pi/10$.  }
  \label{FigVNEcurr[0]}%
\end{figure}

\begin{figure}[h]
  \centerline{\includegraphics[width=9cm,clip]%
    {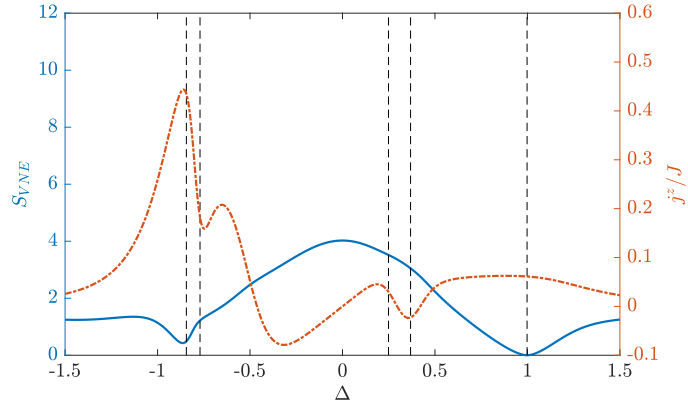}}
  \caption{As in Fig.~\ref{FigVNEcurr[0]} for $\Gamma=10$.  }
  \label{FigVNEcurr[1]}%
\end{figure}

We stress that the spin-helix state is realized as a NESS of the
system only in the ideal limit $\Gamma\to\infty$.  To see how the
above quantities change in the physically more relevant situation of
$\Gamma$ finite, consider first a simple yet demonstrative example.
Figures~\ref{FigVNEcurr[0]} and \ref{FigVNEcurr[1]} show the von
Neumann entropy of the actual $\rho_\mathrm{NESS}$ and the
corresponding steady-state magnetization current $j^z$ as a function
of the anisotropy $\Delta$ for fixed $N$, $\theta$ and $\Phi$, for
two, large and small values of the dissipation strength $\Gamma$ (for
a quantification of the notions ``large'' and ``small'' see
Sec.~\ref{sec::Convergence to the spin-helix state}).  The NESS is
found solving numerically Eq.~(\ref{LME}).

For large $\Gamma$, in Fig.~\ref{FigVNEcurr[0]} we see that for values
of the anisotropy $\Delta$ given by
\begin{align}
  \Delta_\mathrm{cr}(m,\Phi)= \cos \frac{\Phi+2 \pi m }{N-1}, \qquad
  m=0,1,\dots, N-2,
  \label{DeltaCritical}%
\end{align}
$\rho_\mathrm{NESS}$ becomes a pure state, namely, a spin-helix state
with winding number $m$. For the same value of $\Gamma$, the
steady-state magnetization current abruptly changes sign and amplitude
in the region $\De \in [-1,1]$ depending on the value of $\sin
\varphi(m)$.  For small $\Gamma$, see Fig.~\ref{FigVNEcurr[1]}, the
above pure-state features fade away for both $S_{VNE}$ and $j^z$.

If the polarizations of the boundary spins of the chain differ
slightly, as in the example shown in Figs.~\ref{FigVNEcurr[0]} and
\ref{FigVNEcurr[1]}, where the boundary angle mismatch is $\Phi
=\pi/10$, one would expect a steady magnetization current proportional
to the bulk gradient $\Phi/(N-1)$.  In fact, naively, one may assume
that neighbouring spins $k,k+1$ are almost collinear in the steady
state.  This scenario indeed happens for small $\Gamma$, and, if
$\Gamma$ is large, for $\Delta$ away from the critical values
(\ref{DeltaCritical}), where the spins arrange in a helix structure
with angle between neighbouring spins $\varphi=(\Phi+2 \pi m)/(N-1)$,
$m=0,1,\dots, N-2$.  On the other hand, at the critical values of
$\Delta$ corresponding to winding numbers $m>0$, a resonance takes
place with a drastic increase of the amplitude of the steady current
$j^z$.  As the system size grows, the magnetization current and the
von Neumann entropy peaks become narrower and steeper.

\begin{figure}[tbp]
  \centerline{\includegraphics[width=9cm,clip]%
    {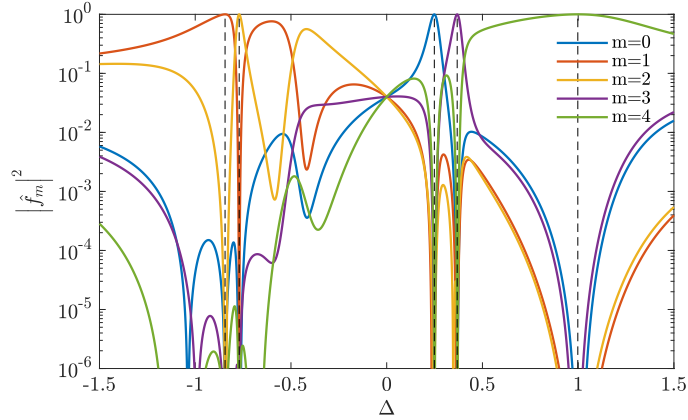}}
  \caption{Generalized Fourier coefficients $\hat{f}_m$ versus the
    anisotropy $\Delta$ for $\Gamma=1000$.  System parameters as in
    Fig.~\ref{FigVNEcurr[0]}.  }
  \label{FigGFT[0]}%
\end{figure}

\begin{figure}[tbp]
  \centerline{\includegraphics[width=9cm,clip]%
    {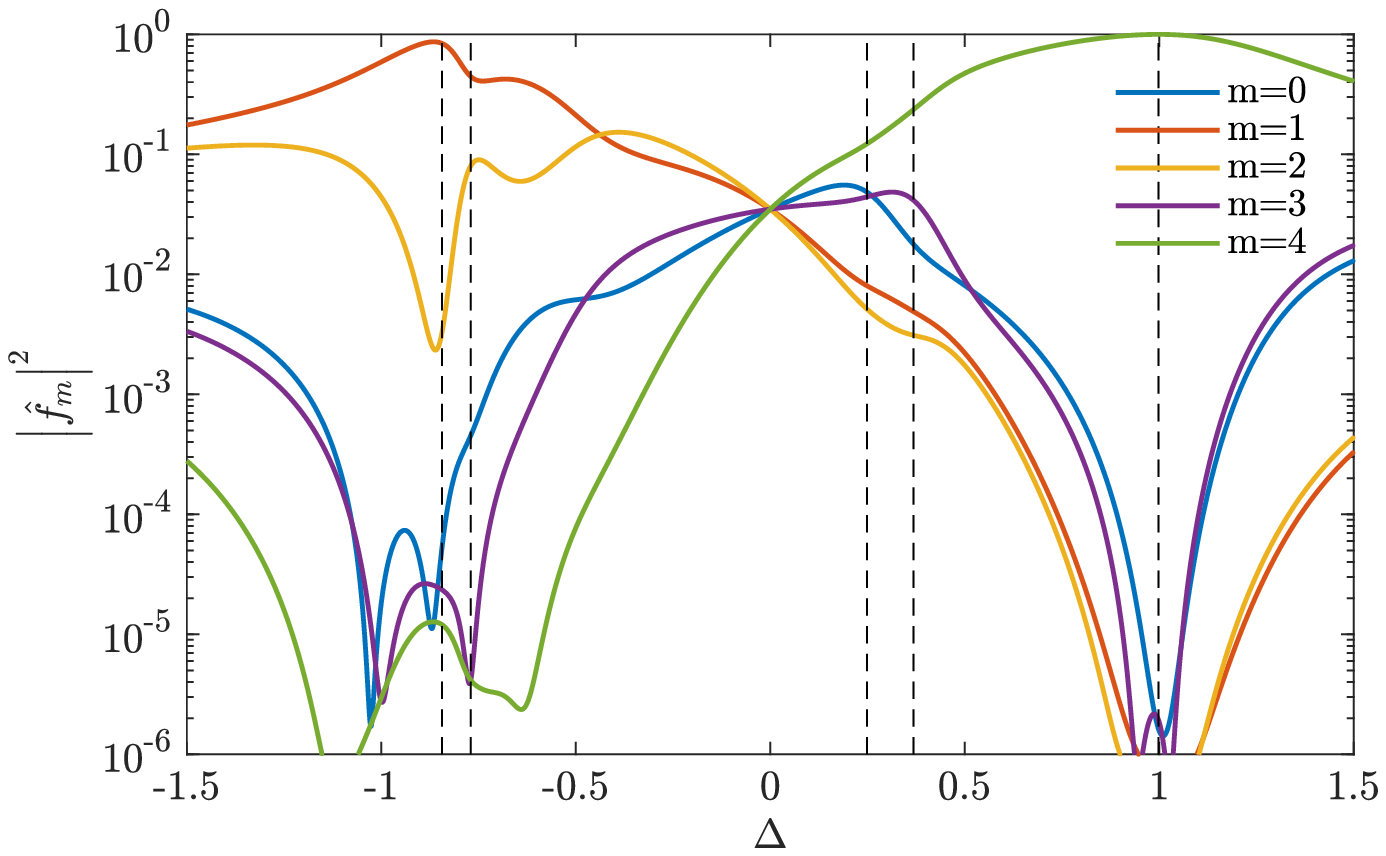}}
  \caption{As in Fig.~\ref{FigGFT[0]} for $\Gamma=10$. }
  \label{FigGFT[1]}%
\end{figure}

To verify the existence of the helix structure of the spins in the
regions near the critical values of the anisotropy, it is instructive
to look at the GFT coefficients $\hat{f}_m$ of
Eq.~(\ref{DefGeneralizedFourierTrasform}) as a function of the
anisotropy. As shown in Figs.~\ref{FigGFT[0]} and \ref{FigGFT[1]}, the
GFT coefficients $\hat{f}_m$ reach their absolute maxima exactly at
the points $\Delta_\mathrm{cr}(m,\Phi)$, in agreement with the
prediction of Eq.~(\ref{SHSfourierCoeff}). This allow us to conclude
that the pure states evidenced in Fig.~\ref{FigVNEcurr[1]} by the
vanishing of $S_{VNE}$ are, in fact, the spin-helix states
(\ref{XXZtargetedstate}).  Note that, for $\Gamma$ large, i.e., in
Fig.~\ref{FigGFT[0]}, the amplitudes of the maxima of the coefficients
$\hat{f}_m$ are independent of $m$ and coincide with the value
$\sin\theta$ predicted by Eq.~~(\ref{SHSfourierCoeff}).

\section{Boundary driven $XXZ$ spin chain: criterion for NESS
  \rev{purity}}
\label{sec::condition for NESS purity}

In order for the NESS to be pure in the Zeno limit, we require an
existence of a NESS expansion in powers of $1/\Ga$,
\begin{align}
  \rho_{\rm{NESS}}(\Gamma) &= \sum_{m=0}^\infty
  \frac{\rho^{(m)}}{\Gamma^m},
  \label{NessExpansion}%
\end{align}
where the zeroth order term is a pure state,
\begin{align}
  \rho^{(0)}&=\ketbra{\Psi}{\Psi}.
  \label{rho0}%
\end{align}
Consistency of the expansion (\ref{NessExpansion}) with the
\rev{purity} assumption (\ref{rho0}) leads to restrictions for the
effective Hamiltonian $H$. The general theory~\cite{ZenoGeneralTheory}
predicts that, for boundary driven systems, in the Zeno limit a pure
steady state $\rho_\mathrm{NESS}=\ketbra{\Psi}{\Psi}$ can be targeted,
where $\ket{\Psi}=\ket{\psi_\mathrm{Zeno}} \otimes
\ket{\psi_\mathrm{target}}$, with $\ket{\psi_\mathrm{Zeno}}\in
\mathcal{H}_0$ and $\ket{\psi_\mathrm{target}}\in \mathcal{H}_1$,
$\mathcal{H}_0$ being the Hilbert subspace where the dissipation
(Lindblad operators) acts and $\mathcal{H}_1$ its complement to the
whole Hilbert space, $\mathcal{H} = \mathcal{H}_0 \otimes
\mathcal{H}_1$.  A necessary condition for this NESS \rev{purity} to
be achieved is that
\begin{equation}
  H  |\Psi \rangle  = \lambda |\Psi \rangle  +
  \kappa \ket{\psi_\mathrm{Zeno}^\perp}  \otimes| \psi_\mathrm{target} \rangle,
  \label{ConditionTargetPure}%
\end{equation}
where $\ket{\psi_\mathrm{Zeno}^\perp}\in\mathcal{H}_0$ is a state
orthogonal to $\ket{\psi_\mathrm{Zeno}}$, $\kappa \neq 0$, and
$\lambda$ is an arbitrary real constant.  The criterion
(\ref{ConditionTargetPure}) gives a necessary condition, while two
extra conditions must be checked to guarantee the convergence of the
NESS to the targeted pure state $\ket{\Psi}$ in Zeno limit. These
extra conditions will be discussed in sec.\ref{sec::divergences of
  Gammach}.

The validity of the \rev{purity} criterion (\ref{ConditionTargetPure})
for the Heisenberg Hamiltonian with the boundary spins $1$ and $N$
attached to polarizing reservoirs, stems from the following property
of the local Hamiltonian density $h_{j,j+1}^{XXZ}(\De)$, with $\De=
\cos \varphi$,
\begin{align}
  &h_{j,j+1}^{XXZ}(\cos \varphi) \ket{\psi(\th,\al)}_{j} \otimes
  \ket{\psi(\th,\al+\varphi)}_{j+1} = -i J \sin\theta \sin\varphi
  \nonumber \\ &\qquad \times \left( \ket{\psi^\perp(\th,\al)}_{j}
    \otimes \ket{\psi(\th,\al+\varphi)}_{j+1} -
    \ket{\psi(\th,\al)}_{j} \otimes
    \ket{\psi^\perp(\th,\al+\varphi)}_{j+1} \right),
\end{align}
where the lower index in a state denotes the respective embedding and
\begin{align}
  &\ket{\psi(\th,\al)} = \binom{\cos \frac{\theta}{2} e^{-i
      \al/2}}{\sin \frac{\theta}{2} e^{i \al/2}},
  \\
  &\ket{\psi^\perp(\th,\al)} = \binom{ \sin \frac{\theta}{2} e^{-i
      \al/2}}{- \cos \frac{\theta}{2} e^{i \al/2}}.
\end{align}
It is simple to verify that Eq.~(\ref{ConditionTargetPure}) is
satisfied with $\la=0,\ka=-i J \sqrt{2} \sin\theta \sin\varphi$, and
\begin{align}
  &\ket \psi_\mathrm{Zeno} = \ket{\psi(\th,0)}_1 \otimes
  \ket{\psi(\th,\Phi)}_N,
  \label{psiZeno}%
  \\
  &\ket \psi_\mathrm{Zeno}^\perp =\frac{1}{\sqrt{2}} \left(
    \ket{\psi^\perp(\th,0)}_1 \otimes \ket{\psi(\th,\Phi)}_N-
    \ket{\psi(\th,0)}_1 \otimes \ket{\psi^\perp(\th,\Phi)}_N \right),
  \label{psiZenoPerp}%
  \\
  &\ket{\psi_\mathrm{target}} = \bigotimes_{j=2}^{N-1}
  \ket{\psi(\th,(j-1)\varphi)}_j.
  \label{psitarget}%
\end{align}
Note that in the above three states we have $\alpha=0$ and
$\varphi=(\Phi+2\pi m)/(N-1)$, with $m$ integer.  We conclude that the
stationary pure state approached in the Zeno limit (fully polarizing
reservoirs), namely,
\begin{equation}
  \ket{\Psi} = \bigotimes_{j=1}^{N} \ket{\psi(\th,(j-1)\varphi)}_j.
  \label{PSI}%
\end{equation}
is the spin-helix state anticipated in Eq.~(\ref{XXZtargetedstate}).
It describes a ``homogeneous'' spin precession around the anisotropy
axis $Z$ along the chain, with a constant polar angle $\theta$ and a
monotonically increasing azimuthal angle $(j-1)\varphi$,
$j=1,2,\dots,N$, matching the boundary values $(\sin\theta
,0,\cos\theta)$ and
$(\sin\theta\cos\Phi,\sin\theta\sin\Phi,\cos\theta)$.  Indeed, the
expectation value of the spin projection at site $j$ is
\begin{align}
  \langle \vec{\sigma}_{j} \rangle &= \tr \left(
    \ketbra{\psi(\th,(j-1)\varphi)}{\psi(\th,(j-1)\varphi)}
    \vec{\sigma} \right) \nonumber \\ &= (\sin\theta\cos
  (j-1)\varphi,\sin\theta\sin (j-1)\varphi,\cos\theta).
\end{align}
For $\theta=\pi/2$, we have the simpler helix-state of
Eq.~(\ref{XYplanetargetedstate}) describing spins which locally rotate
entirely in the $XY$ plane.

Whereas the criterion (\ref{ConditionTargetPure}) ensures that the
NESS converges to the spin-helix pure state in the limit
$\Gamma\to\infty$, the general theory developed in
\cite{ZenoGeneralTheory} allows us to quantify the speed of this
convergence by establishing a characteristic dissipation as discussed
in the next Section.

\section{Convergence to the spin-helix state for finite dissipation
  strength}
\label{sec::Convergence to the spin-helix state}

According to \cite{ZenoGeneralTheory}, we introduce an orthonormal
basis $\ket{e^{j}}$ in the subspace ${\cal H}_0$ where dissipation
acts and split the Hamiltonian $H$ with respect to this basis.  In our
present case ${\cal H}_0$ is the direct product of the local spin
spaces corresponding to the sites $1$ and $N$,
\begin{equation}
  H = \sum_{j=0}^{d_0-1} \sum_{k=0}^{d_0-1} H_{jk}=
  \sum_{j=0}^{d_0-1} \sum_{k=0}^{d_0-1}
  |e^{j}\rangle \langle e^k | \otimes h^{jk},
  \label{Hdecomposition}%
\end{equation}
where $d_0=4$ is the dimension of ${\cal H}_0$.  The first two basis
vectors are chosen as $\ket{e^{0}}\equiv \ket{\psi_\mathrm{Zeno}}$ and
$\ket{e^{1}}\equiv \ket{\psi_\mathrm{Zeno}^\perp}$, with
$\ket{\psi_\mathrm{Zeno}}$ and $\ket{\psi_\mathrm{Zeno}^\perp}$
defined as in Eqs.~(\ref{psiZeno}) and (\ref{psiZenoPerp}),
respectively. The other vectors of the basis are chosen as (lower
indices denote the embedding)
\begin{align}
  \ket{e^{2}} &=\frac{1}{\sqrt{2}} \left( \ket{\psi^\perp(\th,0)}_1
    \ket{\psi(\th,\Phi)}_N + \ket{\psi(\th,0)}_1
    \ket{\psi^\perp(\th,\Phi)}_N \right),
  \label{e2}%
  \\
  \ket {e^{3}} &= \ket{\psi^\perp(\th,0)}_1
  \ket{\psi^\perp(\th,\Phi)}_N.
  \label{e3}%
\end{align}
Having defined the basis in the ${\cal H}_0$, the coefficients
$h^{jk}$ of the decomposition (\ref{Hdecomposition}), which are
operators in $\mathcal{H}_1$, are readily calculated as
\begin{equation}
  h^{jk}= \trb\left(
    \left(\ket{e^k}\bra{e^j}  I_{2,3,\dots,N-1}\right) H \right),
  \label{Defhjk}%
\end{equation}
where $I_{2,3,\dots,N-1}$ is the identity operator in the space of
spins $2,3,\dots,N-1$, and
\begin{align}
  \trb({\cdot}) &= \trn{N}(\trn{1}(\cdot)),
  \\
  \trn{n}(\cdot) &= \left( \bra{+}\cdot\ket{+} \right)_n + \left(
    \bra{-}\cdot\ket{-} \right)_n,\qquad n=1,2,\dots,N,
\end{align}
where $\ket{\pm}$ are the eigenstates of $\sigma^z$.

Of special importance is the term $h^{00}$, which is the projection of
the Hamiltonian $H$ on the state $\ket{e^{0}}$ targeted by the
dissipation. In fact, we have $\mathcal{D}_1
\ket{e^0}\bra{e^0}=\mathcal{D}_N \ket{e^0}\bra{e^0}=0$, where
$\mathcal{D}_1$ and $\mathcal{D}_N$ are the dissipators associated to
the Lindblad operators $L_1$ and $L_N$, e.g., $\mathcal{D}_1 \rho =
L_1 \rho L_1^\dagger -\frac{1}{2} (L_1^\dagger L_1 \rho + \rho
L_1^\dagger L_1)$.  From Eq.~(\ref{Defhjk}) and after some algebra, we
obtain
\begin{align}
  &h^{00} = H'+ C_{++}(2,\th,0) + C_{++}(N-1,\th,\Phi),
  \label{Defh00}%
  \\
  &H' = \sum_{j=2}^{N-2}h^{XXZ}_{j,j+1}(\De),
  \label{DefHprime}%
  \\
  &C_{++}(m,\th,\varphi) = \trn{m-1}\left( \left(\ket{\psi(\varphi)}
      \bra{\psi(\varphi)} \right)_{m-1} h^{XXZ}_{{m-1},m} \right)
  \nonumber \\
  &\phantom{C_{++}(m,\th,\varphi)} = J( \sin \th (e^{i \varphi}
  \si^{-}_m+ e^{-i \varphi} \si^{+}_m) + \De \si^{z}_m\cos \th - \De
  I_{2,3,\dots,N-1} ),
  \label{DefC++}%
\end{align}
where $h^{XXZ}_{j,j+1}$ are the local energy densities of the $XXZ$
Hamiltonian (\ref{DefXXZ}) and $\si^\al_m= I_{2,3,\dots,m-1}\otimes
\si^\al \otimes I_{m+1,\dots,N-1}$, with $1\leq m \leq N$ and
$\al=\pm,z$.  Provided Eq.~(\ref{ConditionTargetPure}) is satisfied,
the target state is an eigenstate of $h^{00}$ with eigenvalue $\la$,
\begin{equation}
  h^{00} \ket{\psi_\mathrm{target}}= \lambda
  \ket{\psi_\mathrm{target}}.
  \label{DefPrincipalEigenvalueLambda}%
\end{equation}

Defining also
\begin{align}
  C_{-+}(m,\th,\varphi) &= \trn{m-1} \left(
    \left(\ket{\psi^\perp(\varphi)} \bra{\psi(\varphi)}\right)_{m-1}
    h^{XXZ}_{m-1,m} \right) \nonumber \\ &= J \left( 2 \sin^2
    \frac{\th}{2} \left( e^{-i \varphi} \si^{+}_m \right) - 2 \cos^2
    \frac{\th}{2} \left( e^{i \varphi} \si^{-}_m \right) + \De
    \si^{z}_m \sin \th \right),
  \label{DefC-+}%
  \\
  C_{+-}(m,\th,\varphi) &= \trn{m-1} \left(
    \left(\ket{\psi(\varphi)}\bra{\psi^\perp(\varphi)}\right)_{m-1}
    h^{XXZ}_{m-1,m} \right) \nonumber \\ &=
  (C_{-+}(m,\th,\varphi))^\dagger,
  \label{DefC+-}%
  \\
  C_{--}(m,\th,\varphi) &= \trn{m-1} \left(
    \left(\ket{\psi^\perp(\varphi)} \bra{\psi^\perp(\varphi)}
    \right)_{m-1} h^{XXZ}_{m-1,m} \right) \nonumber \\ &= J \left(
    -\sin \th \left( e^{i \varphi} \si^{-}_m+ e^{-i \varphi} \si^{+}_m
    \right) -\De \cos \th \si^{z}_m - \De I_{2,3,\dots,N-1} \right),
  \label{DefC--}%
\end{align}
and denoting $\varphi_L=0$ and $\varphi_R=\Phi$, we obtain the
following other components $h^{jk}$:
\begin{align}
  h^{01}&=\frac{1}{\sqrt{2}} \left( C_{-+}(2,\th,\varphi_L) -
    C_{-+}(N-1,\th,\varphi_R) \right),
  \label{Defh01}%
  \\
  h^{02}&=\frac{1}{\sqrt{2}} \left( C_{-+}(2,\th,\varphi_L) +
    C_{-+}(N-1,\th,\varphi_R) \right),
  \label{Defh02}%
  \\
  h^{03}&=0, \qquad \mbox{for $N>2$},
  \label{Defh03}%
  \\
  h^{21}&=\frac{1}{2} \left( C_{++}(N-1,\th,\varphi_R) -
    C_{--}(N-1,\th,\varphi_R) \right.  \nonumber \\ & \qquad \left. +
    C_{--}(2,\th,\varphi_L) - C_{++}(2,\th,\varphi_L) \right),
  \label{Defh21}%
  \\
  h^{31}&=\frac{1}{\sqrt{2}} \left( -C_{+-}(2,\th,\varphi_L) +
    C_{+-}(N-1,\th,\varphi_R) \right),
  \label{Defh31}%
  \\
  h^{11}&=H'-2 \Delta I_{2,3,\dots,N-1},
  \label{Defh11}%
  \\
  h^{jk}&=(h^{kj})^\dagger.
  \label{DefhHermConj}%
\end{align}
The remaining $h^{jk}$ are obtained analogously.

On using Eqs.~(\ref{Defh00}) and (\ref{Defh01}-\ref{Defh31}), we can
compute the characteristic value of the dissipation $\Gamma$, beyond
which the NESS differs from the pure spin-helix state (\ref{PSI}) for
less than a chosen error.  In fact, according to
\cite{ZenoGeneralTheory}, if the criterion (\ref{ConditionTargetPure})
is satisfied, then, not only $\lim_{\Gamma\to\infty}
\rho_\mathrm{NESS} (\Gamma) = \ketbra{\Psi}{\Psi}$, where $\ket{\Psi}$
is given by Eq.~(\ref{PSI}), but for finite $\Gamma$ we have that the
NESS is characterized by a \rev{purity} index
\begin{equation}
  \epsilon (\Gamma)= 1- \tr \rho_\mathrm{NESS}(\Gamma)^2 =
  \frac{\Gamma_\mathrm{ch}^2}{\Gamma^2}+
  o\left( \frac{1}{\Gamma^2} \right),
\end{equation}
whose value is determined by the squared ratio between $\Gamma$ and a
characteristic dissipation. The latter is given by the formula
\begin{equation}
  \left( \Gamma_\mathrm{ch}  \right)^2 =8 |\kappa|^2
  \sum_{\al=1}^{d_1-1}\sum_{\be=1}^{d_1-1}
  (K^{-1})_{\al\be} R_\be,
  \label{ResGammaCharacteristic}%
\end{equation}
where $K$ is the $(d_1-1) \times (d_1-1)$, $d_1= 2^{N-2}$, matrix
having elements
\begin{equation}
  K_{\alpha \beta} = \sum_{k=1}^{d_0-1}
  \left( \left| \langle \alpha | h^{k0} | \beta \rangle \right|^2 -
    \delta_{\alpha \beta}
    \langle \alpha |  (h^{k0})^\dagger h^{k0} | \alpha \rangle \right),
  \qquad \al,\be=1,2,\dots,d_1-1,
  \label{DefK}%
\end{equation}
and
\begin{equation}
  R_\alpha = \bra{\al} F \ketbra{\psi_\mathrm{target}}
  {\psi_\mathrm{target}}F^{\dagger} \ket{\al}, \qquad \al=1,2,\dots,d_1-1.
  \label{defRalpha}%
\end{equation}
with
\begin{equation}
  F = \sum_{k=1}^{d_0-1}
  \left( h^{k1} + \left[\Lambda h^{01}, h^{k0}\right] \right),
  \label{DefF}%
\end{equation}
and
\begin{equation}
  \Lambda = \sum_{\alpha=1}^{d_1-1}
  \frac{1} {\lambda_\alpha-\lambda} | \alpha \rangle \langle \alpha |.
  \label{DefLambda}%
\end{equation}
The symbols $\lambda_\alpha$ indicate the eigenvalues of $h^{00}$ for
$\al=1,\dots,d_1-1$, whereas $\lambda\equiv\la_0$, also entering the
condition (\ref{ConditionTargetPure}), is the principal eigenvalue of
$h^{00}$, see Eq.~(\ref{DefPrincipalEigenvalueLambda}).

In agreement with the results of the previous section, for
$\Delta=\cos\varphi$ and $\varphi=(\Phi+2\pi p)/(N-1)$, with $p$
integer, we find
\begin{equation}
  \la =0,
\end{equation}
which follows from the easily checkable identity
\begin{align}
  C_{++}(m,\th,\Phi) \ket{\psi(\theta,\Phi+\vfi)} =&\ (\cos \vfi -
  \De)
  \ket{\psi(\theta,\Phi+\vfi)} \nonumber \\
  &+ \left( i \sin \vfi \sin \th - \frac{1}{2} \sin 2 \th ~(\cos \vfi
    - \De) \right)\ket{\psi^\perp(\theta,\Phi+\vfi)}
\end{align}
and Eqs.~(\ref{DefPrincipalEigenvalueLambda}) and (\ref{Defh00}).

\section{Divergences of the characteristic dissipation}
\label{sec::divergences of Gammach}
If $\Lambda$ and $K^{-1}$ are nonsingular matrices, the characteristic
value of the dissipation $\Gamma_\mathrm{ch} $ is always finite, and
the NESS converges to the SHS (\ref{XXZtargetedstate}) for $\Gamma \gg
\Gamma_\mathrm{ch}$.  On the other hand, divergence of
$\Gamma_\mathrm{ch} $ signalizes a breakdown of the \rev{purity}
assumption (\ref{rho0}), and consequently the breakdown of the
convergence to the SHS in the Zeno limit.

Points of divergence of $\Gamma_\mathrm{ch}$ may happen when $\Lambda$
or $K^{-1}$ is singular.  In our problem, we have three parameters:
the twisting angle $\varphi$, the polar angle $\th$, and the size of
the system $N$, the anisotropy being fixed by
Eq.~(\ref{ConditionTargetPure}) to be $\Delta=\cos \varphi$.
Investigating, with the help of Mathematica, the analytic expression
(\ref{ResGammaCharacteristic}) for $N\leq 12$ leads us to
formulate the following ansatz.\\
\textit{Ansatz.} For any finite size $N\geq 3$ and any fixed
$0<\theta<\pi$, $\Gamma_\mathrm{ch}(\varphi)$ diverges at a set of
isolated singular points $\varphi^{*}_j \in \Omega_N$, given by
\begin{equation}
  \Omega_N =
  \left\{
    \varphi^{*}_{j}:~ \varphi^{*}_{j} k= \pi d,~ k=2,3,\dots,N-1,~
    d\in\mathbb{Z}
  \right\}.
  \label{ConditionNESSdivergence}%
\end{equation}
Moreover, in the $\epsilon$--vicinity of every point
$\varphi^{*}_{j}$, we have $\Gamma_\mathrm{ch}
(\varphi^{*}_{j}+\epsilon) = A_0(N,\theta,\varphi_j^{*})
|\epsilon|^{-a_j}$, with $a_j=1$.

\subsection{Effective number of $\Gamma_\mathrm{ch}$ singularities}
The condition (\ref{ConditionNESSdivergence}) has a very simple
geometrical interpretation: it marks all possible twisting angles
$\varphi$ for which the target spin-helix state has two or more
collinear spins including the boundary spins.  It is enough to
describe the points of $\Omega_N$ lying in the segment $]0,\pi[$. In
fact, inverting the sign of a $\varphi^*_j$ changes the sign of the
helicity but conserves the collinearity of the spins. We also exclude
the trivial points $\varphi_{j}^*=0,\pi$.  Let us denote $\Omega_N^*$
the reduced set of different values $\varphi^*_j \in ]0,\pi[$.  For
fixed $N$, this set consists of the angles
\begin{align}
  \{\varphi_{j}^* \} = \left\{
    \frac{\pi}{2},~\frac{\pi}{3},\dots,\frac{\pi}{N-1} \right\}
\end{align}
and all different multiples of them such that $0< \varphi_{j}^* d <
\pi$.  For instance, for $N=6$ we explicitly have
\begin{equation}
  \Omega^*_6= \left\{
    \frac{\pi}{2},\frac{\pi}{3},\frac{\pi}{4},\frac{\pi}{5},
    \frac{2\pi}{3},\frac{2\pi}{5},\frac{3\pi}{4},\frac{3\pi}{5},
    \frac{4\pi}{5}
  \right\}.
  \label{ResOmega6}%
\end{equation}
In general,
\begin{align}
  \Omega^{*}_N = \left\{ \varphi^{*}_{j}:~ \varphi^{*}_{j} k = \pi d,~
    k=1,2,\dots,N-1,~d=1,2,\dots,k-1 \right\},
  \label{ResSubsetOmega*}%
\end{align}
with the condition that pairs $d,k$ having the same ratio $d/k$ are
counted only once.

For fixed $N$, the total number of the points where
$\Gamma_\mathrm{ch}$ diverges is given by
\begin{equation}
  \left| \Omega^*_N \right| = \left(\sum_{k=2}^{N-1}
    \sum_{d(k)=1}^{k-1} 1 \right)',
\end{equation}
where the prime means that different pairs $d,k$ with the same ratio
$d/k$ are taken into account one time only in the sum.  For
$N=3,4,5,6,7,8,9,10,100,300$ we find, respectively, $|\Omega^*_N| =
1,3,5,9,11,17,21,27,3003,27317$, where the last two examples have been
computed numerically.  Finding a recursive relation for $\left|
  \Omega^*_N \right|$ is not easy.  Note that if $N_1\equiv
N_\mathrm{pr}$ is a prime number, then $\Omega^*_{N_\mathrm{pr}+1}$
will contain $N_\mathrm{pr}-1$ new elements with respect to
$\Omega^*_{N_\mathrm{pr}}$, namely
\begin{align}
  \Omega^*_{N_{1}+1} \setminus \Omega^*_{N_{1}}&= \left\{ \frac
    {\pi}{N_{1}},\frac {2\pi}{N_{1}}, \dots, \frac
    {(N_{1}-1)\pi}{N_{1}} \right\},
  \label{ResPrimeN}%
\end{align}
so that $|\Omega^*_{N_\mathrm{pr}+1}|= |\Omega^*_{N_\mathrm{pr}}|
+N_\mathrm{pr} -1$. If $N_1$ is not a prime number, then
$|\Omega^*_{N_1+1}|- |\Omega^*_{N_1}| < N_1-1$ since some elements of
the set (\ref{ResPrimeN}) are already present in $\Omega^*_{N_1}$.
Therefore, to find an exact asymptotic behaviour of $|\Omega^*_N|$,
one needs at least to know the distribution of the prime numbers on an
interval $[1,N]$, which is a famous unresolved mathematical
problem~\cite{Ingham}. By using Mathematica, we find that for $N \leq
2000$, the cardinality of $\Omega^*_N$ grows quadratically with the
system size $N$, namely, $|\Omega^*_N| \approx 0.30386 N^2$.

\subsection{NESS at $\Gamma_\mathrm{ch}$ singularities}
\begin{figure}[tbp]
  \centerline{\includegraphics[width=9cm,clip]%
    {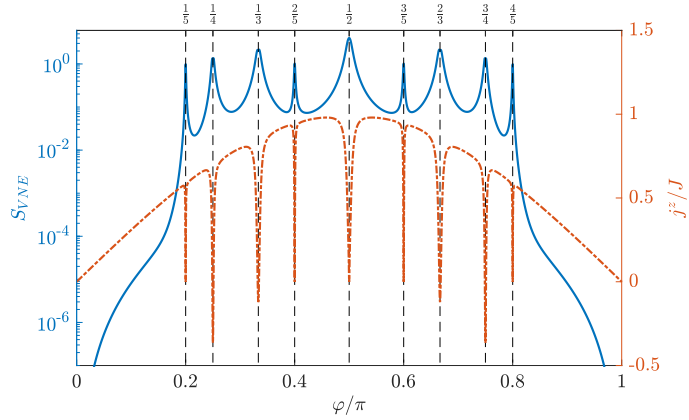}}
  \caption{ Von Neumann entropy of the NESS (solid blue line\rev{,
      left vertical scale}) and steady-state magnetization current
    (dot-dashed red line\rev{, right vertical scale}) as a function of
    the twisting angle $\varphi$. System parameters: $N=6$,
    $\Delta=\cos \varphi$, $\theta=\pi/2$, $\Gamma=500$.  There are
    $9$ singular points characterized by peaks of $S_{VNE}$ and nadirs
    of $j^z$, where the convergence of the NESS to a pure spin-helix
    state fails. These points coincide with those predicted
    theoretically, see Eq.~(\ref{ResOmega6}). Note the symmetry around
    $\varphi=\pi/2$. }
  \label{FigVNEversusC}%
\end{figure}

\begin{figure}[tbp]
  \centerline{\includegraphics[width=9cm,clip]%
    {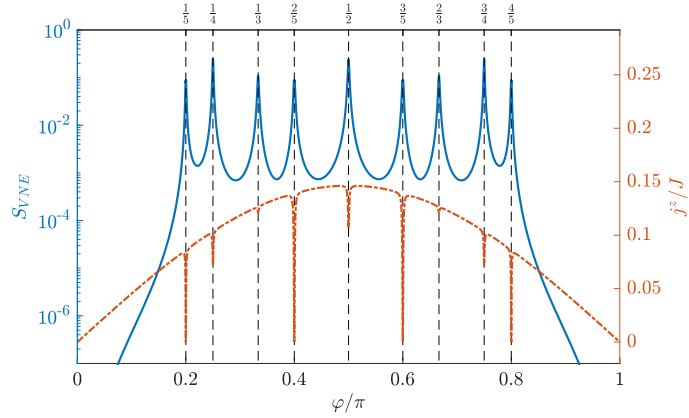}}
  \caption{ As in Fig. \ref{FigVNEversusC} but for $\theta=\pi/8$. }
  \label{FigVNEversusCthetaPi8}%
\end{figure}

On varying the twisting angle $\varphi$ (the anisotropy being fixed at
the value $\Delta=\cos\varphi$), the NESS everywhere converges, in the
Zeno limit, to the pure spin-helix state~(\ref{XXZtargetedstate}),
except for $\varphi$ given by the singular points
(\ref{ConditionNESSdivergence}), where the limiting NESS is mixed.
This is well illustrated by Fig.~\ref{FigVNEversusC}, where the von
Neumann entropy $S_{VNE}(\varphi)$ tends to vanish everywhere except
at the 9 values of $\varphi$ given by Eq.~(\ref{ResOmega6}).  For
small polar angles $\theta$, the convergence of the NESS to the
spin-helix state is faster, see Fig.~\ref{FigVNEversusCthetaPi8}, but
the divergences of $\Gamma_\mathrm{ch}$, where the convergence fails,
arise at the same points.  For the particular value
$\varphi_j^*=\pi/2$, and $\theta=\pi/2$, the NESS can be shown to be a
completely mixed state of the form
\begin{equation}
  \lim_{\Gamma \rightarrow \infty } \rho_\mathrm{NESS} =
  \frac{1}{2^{N-2}} \rho_L \otimes I_{2,3,\dots,N-1} \otimes\rho_R,
  \qquad \mbox{for $\varphi=\pi/2$},
  \label{case.phi=pi/2}%
\end{equation}
where $\rho_L ,\rho_R $ are the reservoir polarizations, see Sec. 4.1
of \cite{XYtwist} for details.  For other $\varphi^*_j$, the NESS
converges to some unknown mixed states.

\subsection{Source of $\Gamma_\mathrm{ch}$ singularities}
All points of divergence of $\Ga_\mathrm{ch}$ must be either due to
divergence of $K^{-1}$, entering Eq.~(\ref{ResGammaCharacteristic})
directly, or the divergence of $\Lambda$ entering the expression
(\ref{ResGammaCharacteristic}) through the terms $Q_{\al,k}=\bra{\al}
[\Lambda h^{01},h^{k0}\ket{0} $, or both \cite{ZenoGeneralTheory}.  By
using Mathematica we checked that each of the points
(\ref{ConditionNESSdivergence}) correspond to a divergence of either
$K^{-1}$ or the $Q_{\al,k}(\Lambda)$ terms.  The partition of the
singularities of $\Gamma_\mathrm{ch}$ between $K^{-1}$ and $\Lambda$
depends quite crucially on the value of the polar angle $\theta$.

For $\theta \neq \pi/2$, we have $\det K\neq 0$ for any $\varphi$ and
any $N$, therefore $K^{-1}$ always exists and all the points of
divergence of $\Ga_\mathrm{ch}$ are due to divergencies of the terms
$Q_{\al,k}(\Lambda)$.

\begin{figure}[tbp]
  \centerline{\includegraphics[width=9cm,clip]%
    {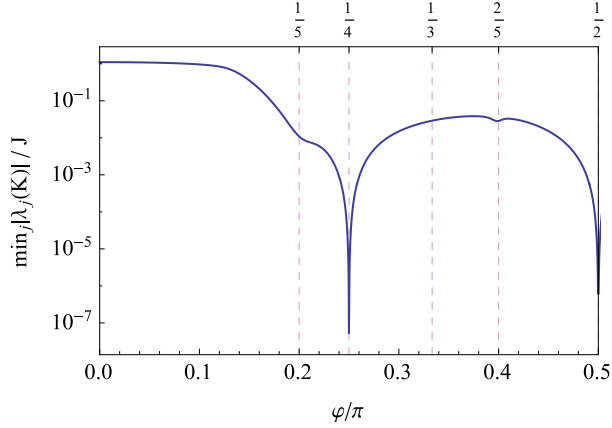}}
  \caption{Minimum modulus of the eigenvalues of of the matrix $K$ as
    a function of the twisting angle $\vfi$ for $N=6$. The plot is
    symmetric with respect to $\varphi=\pi/2$ and $\varphi=0$.
    Parameters: $\theta=\pi/2$, $\Delta=\cos\varphi$.  }
  \label{FigGapK}%
\end{figure}

For $\theta=\pi/2$, $K^{-1}$ is singular at the isolated points
$\varphi^{**}_j \in \Omega^{(K)}_N \subset \Omega^*_N$, where
\begin{align}
  \Omega^{(K)}_N = \left\{ \varphi^{**}_{j}:~ \varphi^{**}_{j} 2k =
    \pi d,~ k=1,2,\dots,\left\lfloor \frac{N-1}{2}
    \right\rfloor,~d=1,2,\dots,2k-1 \right\}.
  \label{ResSubsetOmega(K)}%
\end{align}
In this set, as in $\Omega^*_N$, pairs $d,k$ with the same ratio
$d/(2k)$ are counted only once.  The terms $Q_{\al,k}(\Lambda)$
diverge at the points of the complementary subset
\begin{align}
  \Omega^{(\Lambda)}_N= \Omega^{*}_N \setminus \Omega^{(K)}_N.
  \label{ResSubsetOmega(Lambda)}%
\end{align}
For example, in the case $N=6$ we have
\begin{equation}
  \Omega^{(K)}_6= \left\{
    \frac{\pi}{2},\frac{\pi}{4},\frac{3\pi}{4}
  \right\},
  \label{ResOmega6(K)}%
\end{equation}
\begin{equation}
  \Omega^{(\Lambda)}_6= \left\{
    \frac{\pi}{3},\frac{\pi}{5},
    \frac{2\pi}{3}, \frac{2\pi}{5}, \frac{3\pi}{5},
    \frac{4\pi}{5}
  \right\}.
  \label{ResOmega6(Lambda)}%
\end{equation}
In Fig.~\ref{FigGapK} we show the minimum modulus of the eigenvalues
of of the matrix $K$ as a function of the twisting angle $\vfi$ for
$N=6$ and $\theta=\pi/2$. Zeros are obtained exactly at the points of
the set (\ref{ResOmega6(K)}).

The number of points in $\Omega^{(\Lambda)}_N$ is smaller than the
number of points where $\Lambda$, tout court, has a divergence,
namely, the points of degeneracy of $h^{00}$. This is easily
understood if, instead of directly studing the divergences of the
terms $Q_{\al,k}(\Lambda)$, we proceed as follows.  A divergence of
$\Ga_\mathrm{ch}$ governed by the $\Lambda$ matrix stems from an
inconsistency of the linear system of equations for the coefficients
$M^{(1)}_{\al 0}$ arising in the first order expansion of the NESS in
powers of $1/\Ga$ \cite{ZenoGeneralTheory}. In the basis in which
$h^{00}$ is diagonal, this system has the form, see Eq.~(A24) of
\cite{ZenoGeneralTheory},
\begin{equation}
  (\la_\al - \la_0) M^{(1)}_{\al 0} =
  2 i \kappa \bra{\al}h^{01} \ket{0},\qquad \al=1,2,\dots,d_1-1.
  \label{DefM1}%
\end{equation}
Since $\kappa \neq 0$, the quantity $M^{(1)}_{\al 0}$ diverges if two
conditions are simultaneously satisfied: (a) the eigenvalue $\la_0$ of
$h^{00}$ is degenerate, i.e., $\la_\al - \la_0=0$ for some
$\al=1,2,\dots,d_1-1$, and (b) for the corresponding $\al$ it results
$\bra{\al}h^{01}\ket{0} \neq 0$.  Note that the sole degeneracy of
$\la_0$ may not lead to a divergence of $\Ga_\mathrm{ch}$.

\begin{figure}[h]
  \centerline{\includegraphics[width=9cm,clip]%
    {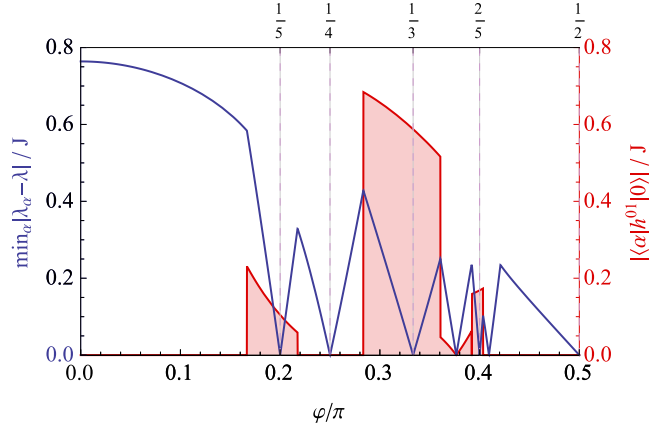}}
  \caption{Gap of the $h^{00}$ spectrum (solid blue line, left
    vertical scale) and corresponding matrix element $\bra{\al} h^{01}
    \ket{0}$ (dashed red line with filling to the horizontal axis,
    right vertical scale) as a function of the twisting angle $\vfi$
    for $N=6$.  Note that the gap vanishes and simultaneously
    $\bra{\al} h^{01} \ket{0} \neq 0$ only at points $\vfi/\pi =
    1/5,1/3,2/5$.  The plot is symmetric with respect to
    $\varphi=\pi/2$ and $\varphi=0$.  Parameters: $\theta=\pi/2$,
    $\Delta=\cos\varphi$. }
  \label{FigH00degeneraciesN6}%
\end{figure}

Inspecting, for various finite $N$, the angles $\varphi$ where both
(a) and (b) conditions are satisfied, we recover the subset
$\Omega^{(\Lambda)}$ given by Eq.~(\ref{ResSubsetOmega(Lambda)}).  The
case $N=6$ with $\theta=\pi/2$ is shown in
Fig.~\ref{FigH00degeneraciesN6}.  Conditions (a) and (b)
simultaneously hold at the points $\varphi/\pi=1/5,1/3,2/5$ as well as
in the symmetric points $\varphi/\pi=3/5,2/3,4/5$ not shown in the
plot.

\begin{figure}[h]
  \centerline{\includegraphics[width=9cm,clip]{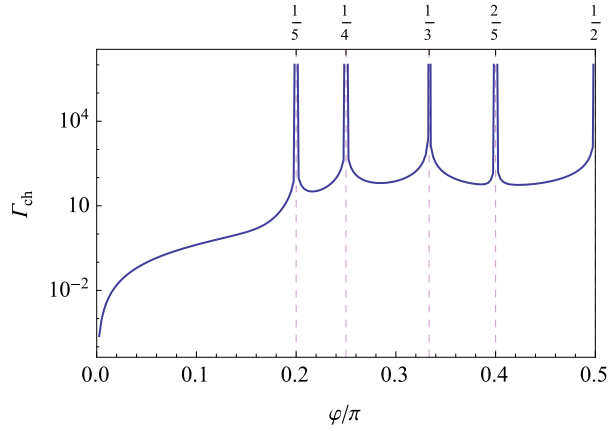}}
  \caption{Characteristic dissipation $\Gamma_\mathrm{ch}$, computed
    according to Eq.~(\ref{ResGammaCharacteristic}), as a function of
    the twisting angle $\vfi$ for $N=6$. The plot is symmetric with
    respect to $\varphi=\pi/2$ and $\varphi=0$.  Parameters:
    $\theta=\pi/2$, $\Delta=\cos\varphi$. }
  \label{FigGammaCritical}%
\end{figure}

We conclude that the NESS reached in the Zeno limit becomes pure for
all $\varphi$, except at the singular points of the set
(\ref{ConditionNESSdivergence}), where two or more spins in the target
spin-helix configuration become collinear.  In
Fig.~\ref{FigGammaCritical} we plot $\Ga_\mathrm{ch}(\varphi)$
evaluated according to Eq.~(\ref{ResGammaCharacteristic}).  The
characteristic dissipation shows divergences exactly at the points
predicted by (\ref{ConditionNESSdivergence}). Note that in the
thermodynamic limit $N\rightarrow \infty$ the number of divergencies
grows quadratically with the system size.

\section{Experimental scenarios}
\label{sec::Experimental scenarios}
Finally, we comment on two hypothetical experimental scenarios.  Using
single atom techniques~\cite{2016OtteXXZ}, it should be possible to
realize systems with a fixed number $N$ of spins $1/2$ coupled via
Heisenberg exchange interaction, and to manipulate either the total
twisting angle $\Phi$ (scenario A), or the anisotropy $\De$ (scenario
B).  For both cases, we assume that the dissipative strength $\Ga$ can
also be controlled. Note that quantum Zeno dynamics \cite{Zeno} is
well within reach of contemporary experimental setups
\cite{ZenoDynamicsExperimental}.

\subsection{Scenario A. Fixed anisotropy $|\Delta|<1$, varying
  boundary twist $\Phi$}

It is clear from the previous discussion that an interesting case
occurs if the anisotropy $\De$ obeys $-1<\De<1$ and $\De\neq 0$.  For
a quantum chain, the regime $-1<\De<1$ is referred to as critical, or,
an easy-plane regime, while the condition $\De\neq 0$ rules out the
so-called noninteracting free-fermion case.  The latter case
corresponds to $\varphi=\pi/2$ and does not converge to a pure NESS
for any $N$, see Eq.~(\ref{case.phi=pi/2}).  Measuring any one of the
NESS properties (a)-(c) discussed in Sec.~\ref{sec::Model}, for
different total boundary twisting angles $\Phi$, we will find a
resonance-like behaviour in correspondence of the existence of a
spin-helix pure NESS. This will happen at the value $\Phi=\Phi_0$
given by
\begin{align}
  \Phi_0 = (N-1) \arccos \Delta .
\end{align}
The value of the characteristic dissipation above which the above
resonance will be measured, can be computed analytically using
Eq.~(\ref{ResGammaCharacteristic}). We have seen that the
characteristic value $\Ga_\mathrm{ch}$ becomes large in proximity of
the singular points $\varphi^*_j$, where a divergence of
$\Ga_\mathrm{ch}$ takes place, see
Eq.~(\ref{ConditionNESSdivergence}). The larger is the size of the
system, the smaller is the distance between two consecutive singular
points $\varphi^*_j$. Therefore, we expect $\Ga_\mathrm{ch}(\Delta)$,
with $\Delta$ corresponding to some generic irrational value of
$\varphi=\arccos \De$, to increase with the system size $N$, and to
diverge in the thermodynamic limit.

\subsection{Scenario B. Fixed boundary twist $\Phi$, varying
  anisotropy $|\Delta|<1$}

This scenario is more spectacular then the previous one.  For every
generic fixed twisting angle $\Phi$, such that $\Phi/\pi$ is an
irrational, there will be $N-1$ resonance values of the anisotropy,
corresponding to the formation of spin-helix states in the Zeno limit.
These resonance values are given by $\De(m)=\cos((\Phi+2 \pi
m)/(N-1))$, $m=0,1,\dots,N-2$.  The characteristic dissipation, above
which the phenomenon can be measured, will depend on $m$ and on the
closeness of the respective $\varphi= \arccos \De(m)$ to the nearest
singular points $\varphi^*_j$ of
Eq.~(\ref{ConditionNESSdivergence}). By increasing the size of the
system we will have two competing effects.  The number of resonances
will grow linearly with the size $N$, but the majority of the
spin-helix states will become more difficult to measure due to the
overall growth of $\Ga_\mathrm{ch}$.  Note, however, that spin-helix
states with the effective smallest winding numbers, namely, $m=0$ and
$m=N-2$, become more accessible as the system size grows. In fact, we
have observed in \cite{PhysRevA.93.022111} that, for these winding
numbers, $\Ga_\mathrm{ch}(\epsilon)$, the characteristic dissipation
at a chosen \rev{purity} $\epsilon$ of the NESS, decreases by
increasing $N$.

\section{Conclusions}
\label{sec::Conclusions}

We have shown that a boundary-driven Heisenberg spin chain in the
critical regime $|\De|<1$ exhibits, for large dissipation strength, a
set of structural transitions in its nonequilibrium steady state
between states with spatially smooth magnetization profile and
spin-helix structures, where the local magnetization significantly
changes from one site to another.  Each spin-helix structure can be
understood as a single generalized discrete Fourier harmonic,
compatible with the boundary conditions imposed by the dissipation.
Varying the anisotropy inside the critical easy-plane phase,
$-1<\De<1$, and keeping sufficiently large dissipation strength, i.e.,
suppressing the boundary fluctuations, a complete set of these
generalized discrete Fourier harmonics can be generated, one by one.

Note that with the help of spin-helix states (\ref{XXZtargetedstate})
one can prepare a single spin in an arbitrary pure state, regardless
of the length of the chain and the position of the spin, for any value
of the anisotropy. This requires manipulating boundary dissipation at
the ends of the chain.  For isotropic spin exchange an arbitrary
single spin state at a distance can be generated with just one
boundary dissipator \cite{2016ZnidarichTargeting}.  From the quantum
transport point of view, existence of spin-helix states allows to
reach ballistic current in a situation where typical current is
diffusive or subdiffusive.

Interestingly, a structural transition to a spin-helix state fails,
whenever two or more spins in the helix become collinear.  Such a
situation takes place when the twisting angle governing the helix is a
rational number of $\pi$ and the spin chain is sufficiently long.  At
a deeper level, these breakups of convergence of the NESS to a pure
state are related to divergences of the characteristic dissipation, a
threshold value of the dissipation, above which the structural
transitions can be observed.  We have provided an explicit formula for
the characteristic dissipation and a detailed classification of its
divergences.

The method we propose can be straightforwardly generalized, and can be
tested on other systems, e.g. on spin chains with higher spins, see
\cite{PopkovSchutzSHS}.  It would be interesting to see if
spin-helix--like structures can be realized in $1D$ arrays of magnetic
atoms ~\cite{2016OtteXXZ}.

\ack Support from DFG grant is gratefully acknowledged. VP thanks the
IBS Center of Theoretical Physics of Complex systems in Daejeon,
Korea, where a part of this work was done, for hospitality.

\appendix

\section{Dependence of $\Ga_\mathrm{ch}$ on the polar angle
  $\theta$. Solvable case $N=3$}
It is instructive to consider the simplest yet nontrivial chain with
$N=3$ spins. In this case, the Hilbert spaces $\mathcal{H}_0$ and
$\mathcal{H}_1$ have dimensions $d_0=2^2$ and $d_1=2^1$, respectively.
The matrix $K$ is, therefore, a scalar and, using Eqs.~(\ref{Defh00}),
(\ref{DefK}), (\ref{Defhjk}), and (\ref{ResGammaCharacteristic}), we
find
\begin{align}
  K = -2 \left(2 \cos (2 \varphi ) \sin ^2(\theta )+\cos (2 \theta
    )+3\right).
\end{align}
Note that $K \leq 0$ for all $\theta,\ \varphi$ and $K=0$ only for
$\theta=\pi/2,\ \varphi=\pi/2$.  For $\Ga_\mathrm{ch}$ we obtain the
remarkably simple expression
\begin{align}
  \Ga_\mathrm{ch}^2 = 8 \sin ^4(\theta ) \sin ^2(\varphi )
  \tan^2(\varphi ).
\end{align}
We conclude that for $N=3$ the dependence of $\Ga_\mathrm{ch}$ on
$\theta$ is described by a multiplicative factor $\sin^2(\theta )$,
which reaches its maximum at $\theta =\pi/2$.

\begin{figure}[tbp]
  \centerline{\includegraphics[width=9cm,clip]%
    {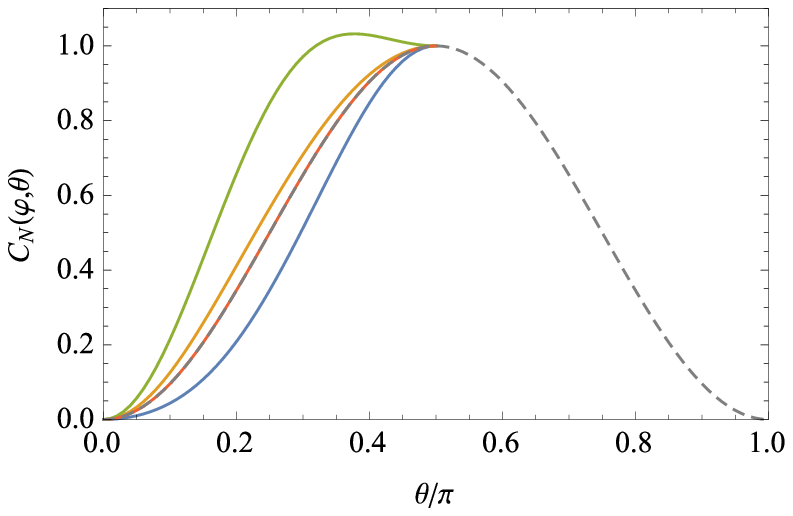}}
  \caption{ Function $C_N(\varphi,\theta)$ versus $\theta$ for $N=5$
    and $\varphi=\pi/2+0.01,2\pi/7,\pi/5,\pi/100$ (solid curves from
    bottom to top).  The dashed gray line, shown for comparison, is
    $C_3(\varphi,\theta)=\sin^2\theta$. }
  \label{FigThetaDependence}%
\end{figure}

For $N>3$, the $\theta$-dependence of $\Ga_\mathrm{ch}$ is no longer
multiplicative, however, it has the form
\begin{align}
  \frac{\Ga_\mathrm{ch}(N,\varphi,\theta)}{\Ga_\mathrm{ch}(N,\varphi,\pi/2)}
  &=   C_N(\varphi,\theta),\\
  C_N(\varphi,\theta) &= C_N(\varphi,\pi-\theta),\\
  C_N(\varphi,\theta) &\leq 1.
\end{align}
We have seen that $C_3(\varphi,\theta) =\sin^2(\theta)$, independent
of $\varphi$.  For $N>3$, the function $C_N(\varphi,\theta)$ at fixed
$\varphi$ is always a symmetric function,
$C_N(\varphi,\theta)=C_N(\varphi,\pi-\theta)$, which has an extremum
at $\theta=\pi/2$.  For most values of $\varphi$,
$C_N(\varphi,\theta)$ has, as a function of $\theta$, an absolute
maximum at $\theta=\pi/2$, see Fig.~\ref{FigThetaDependence}. For
small $\theta \ll 1$, $C_N(\varphi,\theta)$ decreases as $ \theta^2$,
making the respective dissipative pure state (\ref{XXZtargetedstate})
easier to reach (given \rev{purity} attained at smaller dissipative
strengths).  This is in accordance with physical intuition since for
small $\theta$ the spin-helix state (\ref{XXZtargetedstate})
corresponds to small deviations of the local magnetization vector from
the $(0,0,1)$ direction, which are easier to sustain.

% \nocite{*}
\section*{References}

% \bibliography{Zeno} % run pdflatex bibtex pdfltex pdflatex
% \input{thisfile.bbl} % to be used before submission

\providecommand{\newblock}{}

\end{document}